\documentclass[a4paper]{article}

\usepackage{INTERSPEECH2019}

\title{Dialect Identification of Spoken North S\'{a}mi Language Varieties Using Prosodic Features}
\name{Sofoklis Kakouros$^1$, Katri Hiovain$^1$, Martti Vainio$^1$, Juraj \v{S}imko$^1$}
\address{
  $^1$Department of Digital Humanities, University of Helsinki, Finland}
\email{\{sofoklis.kakouros,katri.hiovain,martti.vainio,juraj.simko\}@helsinki.fi}

\begin{document}

\maketitle
\begin{abstract}
  This work explores the application of various supervised classification approaches using prosodic information for the identification of spoken North S\'{a}mi language varieties. Dialects are language varieties that enclose characteristics specific for a given region or community. These characteristics reflect segmental and suprasegmental (prosodic) differences but also high-level properties such as lexical and morphosyntactic. One aspect that is of particular interest and that has not been studied extensively is how the differences in prosody may underpin the potential differences among different dialects. To address this, this work focuses on investigating the standard acoustic prosodic features of energy, fundamental frequency, spectral tilt, duration, and their combinations, using sequential and context-independent supervised classification methods, and evaluated separately over two different units in speech: words and syllables. The primary aim of this work is to gain a better understanding on the role of prosody in identifying among the different language varieties. Our results show that prosodic information holds an important role in distinguishing between the five areal varieties of North S\'{a}mi where the inclusion of contextual information for all acoustic prosodic features is critical for the identification of dialects for words and syllables.
\end{abstract}

\section{Introduction}

Dialects can be generally defined as language varieties that characterize the form of the language within a specific, typically geographic, region or community (see, e.g., \cite{sumner2009effect}). The perception of dialectal cues in speech is particularly important as it has wide ranging implications in shaping attitudes and signalling the origin of the speaker \cite{preston2005dialect}. Overall, the task of identifying a dialect is similar to the more general problem of language identification, where, however, distinguishing among dialectal varieties introduces more challenges as it involves subtler differences between variants of the same language.

Previous studies have indicated that the perception of language differences seems to be dependent on a complex interplay of segmental (phonemes), suprasegmental (prosody, phonotactics), and higher-level information (such as lexemes) (see, e.g., \cite{rouas2007automatic,barkat2001perceptual,vicenik2013role}). The same information seems to be relevant also for dialect identification (see, e.g., \cite{vicenik2013role,clopper2004some,clopper2007native}). For instance, it has been shown that adults can distinguish between languages that are prosodically similar by making use of either rhythmic timing or pitch information \cite{vicenik2013role}.

In automatic language identification (LID), most methods are based on the processing of lexical, phonotactic, and acoustic features (see, e.g., \cite{khurana2017qmdis}). In general, the task of a LID system is to classify a given spoken utterance into one out of many languages. Similarly, for automatic dialect identification (DID), the task is to classify an utterance among many spoken dialects within a language. Overall, DID has remained relatively unexplored compared to LID, perhaps due to the inherent challenges in DID (dialectal similarities tend to be much higher compared to languages) but also due to the limited data resources available.

Most systems for DID make use of an array of approaches that can be categorized into the following: (i) lexical, (ii) phonotactic, and (iii) acoustic. Lexical and phonotactic techniques are typically based on extracting word or phone representations using, for instance, \textit{n}-gram statistics (see, e.g., \cite{khurana2017qmdis, najafian2018exploiting, soufifar2011ivector, ali2015automatic}). Acoustic approaches utilize acoustic features, such as Mel Frequency Cepstral Coefficients (MFCCs), that are then typically modelled with an i-Vector framework (see, e.g., \cite{zhang2018language, trong2016deep}). Currently, the state-of-the-art methods in DID and LID are based on using deep bottleneck features (BNF) in order to extract frame-level features for i-Vector systems (see, e.g., \cite{trong2016deep}).

Overall, in the context of LID and DID, prosody has not been studied extensively, perhaps due to the inherent challenges in modelling the dynamic nature of prosodic phenomena that entail temporal variations across long sequences in speech. To address this, in this work we examine the prosodic differences between the five language varieties of North S\'{a}mi and determine how important is the role of prosody in distinguishing among the five dialects. Our target is not to improve over existing state-of-the-art DID systems, but rather to gain an understanding of the importance and relevance of prosody in dialect identification but also to potentially identify important prosodic factors to consider in the development of future systems. Next, the North S\'{a}mi language and its dialects are briefly presented.

\subsection{The North S\'{a}mi language and dialects}

The North S\'{a}mi language is an indigenous and endangered language with speakers covering a geographic area spanning the northern regions of three countries: Finland, Norway, and Sweden. North S\'{a}mi is the most widely spoken among the S\'{a}mi languages with approximately 25 000 speakers \cite{aikio2015variation,sammallahti1998saami}. As a minority language, practically all North S\'{a}mi speakers are bilingual in S\'{a}mi and the majority language of the country they live in. However, North S\'{a}mi is a majority language in only two regions, in Kautokeino and Karasjoki in Norway \cite{aikio2015variation}. 

North S\'{a}mi belongs to the Uralic language family and is related to, for instance, Finnish, Estonian, and Hungarian. There are 10 S\'{a}mi languages altogether \cite{kulonen2005saami} that differ from each other in many linguistic aspects such as phonology, morphology, syntax, and lexicon. In general, adjacent S\'{a}mi languages can be mutually intelligible.

The North S\'{a}mi language is further divided into three main dialect groups: Sea S\'{a}mi (northernmost coastal area in Norway), Finnmark S\'{a}mi (northern Finland and Norway), and Torne S\'{a}mi (west from Finnmark S\'{a}mi speaking area, and northernmost parts of Sweden) \cite{sammallahti1998saami}. The data used in this work represents the Finnmark North S\'{a}mi variety. The Finnmark North S\'{a}mi dialect group is still traditionally divided into Western and Eastern subdialects \cite{sammallahti1998saami}. The speech data used for this work represents both of these subdialects and speakers from both Norway and Finland. This means that in addition to the dialectal differences (mostly concerning phonetic and [morpho]phonological features), there are two majority languages, Finnish and Norwegian, which are in constant language contact with the S\'{a}mi languages \cite{aikio2015variation}.

In previous works on North S\'{a}mi language varieties, differences between the five dialects have been identified using i-vectors \cite{jokinen2016variation} and evidence of their typological differences have been observed \cite{hiovain2018mapping}. In the present study we investigate the potential prosodic differences among the five regional varieties of North S\'{a}mi and explore whether the geographical dispersion of North S\'{a}mi also underpins disparities in the production of North S\'{a}mi that are reflected in the acoustic prosodic characteristics of North S\'{a}mi speakers. Moreover, we investigate the features and units of analysis that hold a role in discriminating between the dialects.

\section{Materials and Methods}

The material used in this study consists of North S\'{a}mi continuous speech from recordings of speakers from five geographically distinct locations. Analysis of the data is based on the extraction of standard acoustic features that are commonly utilized for speech analysis (see, e.g., \cite{kakouros20163pro}).

\subsection{The extended DigiSami corpus}
\label{sec:extendedDigisamiCorpus}

\begin{table}[t]
  \centering
  \caption{Number of speakers, number of recordings, and overall duration of recordings for each variety in the extended DigiSami corpus. Light gray denotes the recordings from Finland and dark gray those from Norway.}
  \includegraphics[width=\linewidth]{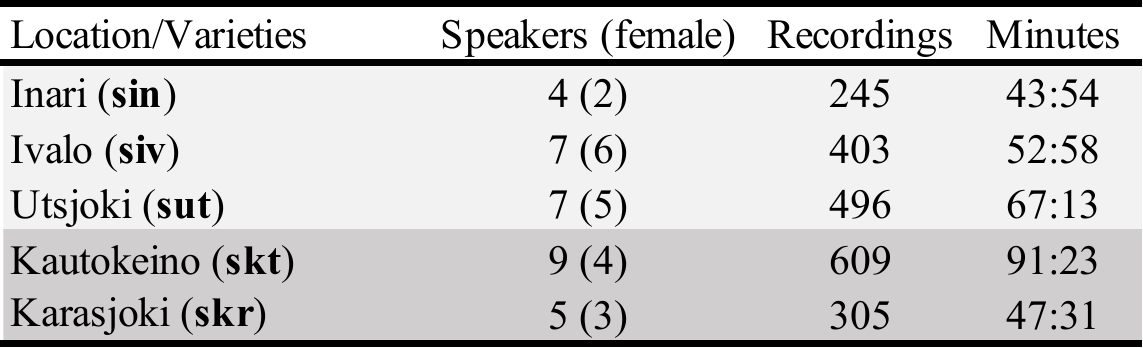}
  \label{tab:speaker_durs}
\end{table}

The DigiSami corpus \cite{jokinen2017digisami} is a collection of read speech recordings from five locations that have been traditionally inhabited by the S\'{a}mi: Inari, Ivalo, and Utsjoki in Northern Finland, and Kautokeino and Karasjoki in Northern Norway. The task of the speakers was to read aloud Wikipedia articles about S\'{a}mi languages and traditional costumes. The corpus consists of speech data from 25 (16 female) native North S\'{a}mi speakers (16-65 years of age) recorded using a 4-channel portable recorder (Roland Edirol R-4 Pro) and condenser mini-lavalier microphones (AKG C 417L). The data also contain annotations at the word, sentence, and phrase level that were produced with a combination of manual annotations and forced alignment (using WebMAUS Basic, see \cite{schiel1999automatic,kisler2017multilingual}). For a more detailed description of the data collection process and corpus, see \cite{jokinen2014open,jokinen2016variation,jokinen2017digisami}.

To extend the original DigiSami corpus \cite{jokinen2017digisami} (due to the limited duration of the corpus recordings), additional recordings were collected from seven native S\'{a}mi speakers (4 female, age range 22-64) at two locations, namely, at Oulu University (Finland) and Kautokeino (Norway) (see Table~\ref{tab:speaker_durs} for an overview). Specifically, two speakers were recorded at Oulu University (originally from Utsjoki and Ivalo), and five speakers at Kautokeino (four originally from Kautokeino and one from Karasjoki). The recording process was the same as in the DigiSami corpus \cite{jokinen2017digisami}: the speakers read aloud the same Wikipedia articles in North S\'{a}mi (with the inclusion of some additional sentences). For the recordings, a Zoom H2n portable recorder with five built-in microphones (using Mid-Side recording mode) was used. The new data were annotated with the same procedure as in the original DigiSami corpus to make the data compatible with the DigiSami corpus (e.g., annotations at the word, sentence, and phrase level). This resulted, in total, to 32 speakers and approximately 5 hours of speech data for the \textit{extended DigiSami corpus}.

In the experiments, a 4-fold evaluation procedure was used where speakers from each dialect where split into four groups to form batches of approximately equal duration in minutes --no speaker occurring at the same time in the training and test sets. Although this resulted in balanced speech data within each dialectal group, this was not the case across dialectal groups where the batch duration was different for each dialect (see Table \ref{tab:speaker_durs}). However, given the current data, this type of data division was deemed optimal for the current experiments. The 4-fold evaluation process was repeated five times (with different training and test sets) and the final results were averaged.

\begin{figure}[t]
  \centering
  \includegraphics[width=\linewidth]{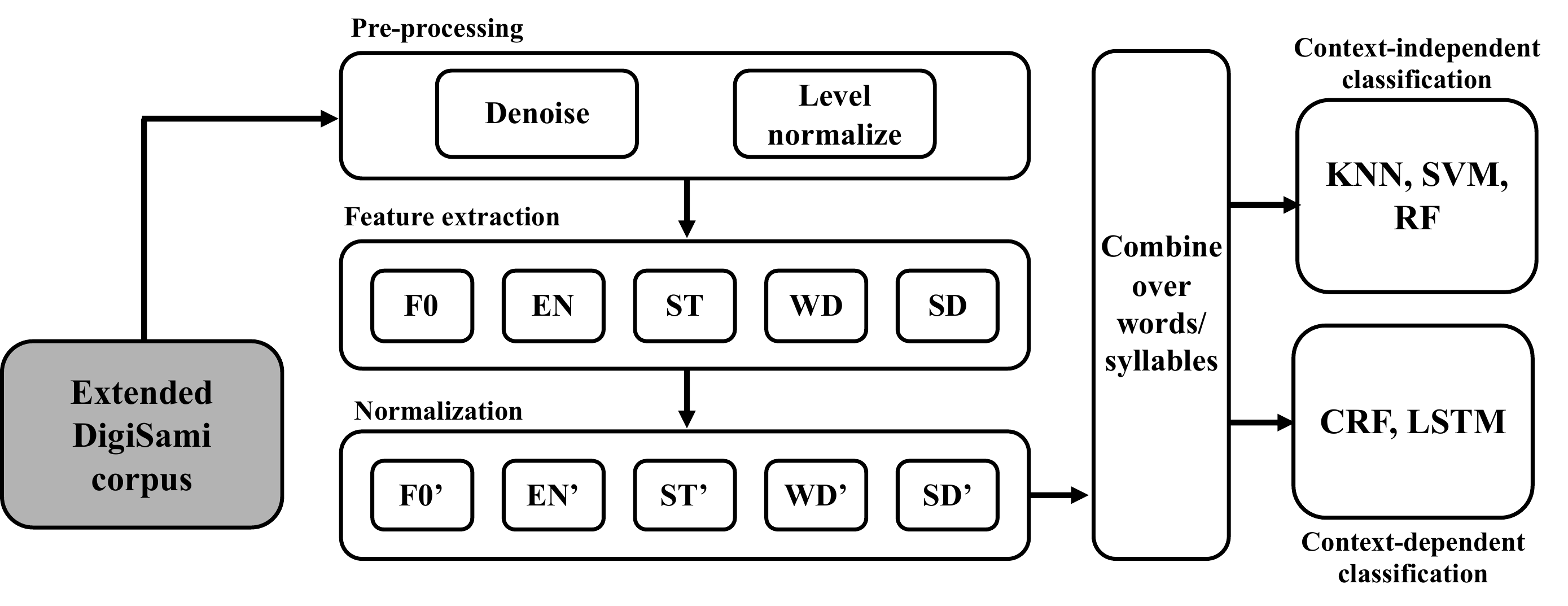}
  \caption{Processing pipeline for the signals.}
  \label{fig:proc_pipeline}
\end{figure}

\subsection{Feature extraction}

Extraction of the acoustic features from the speech signals first involved a pre-processing stage that was followed by computation of the features and normalization.

\subsubsection{Pre-processing}

As majority of the speech recordings were carried out in non-ideal conditions, thus introducing varying types of noise as well as irregular sound levels across the recordings, all signals were first passed through a pre-processing pipeline. Following manual inspection of the signals it was observed that the noise component present within individual recordings was relatively stationary. Therefore, spectral noise subtraction was utilized to denoise the signals \cite{boll1979suppression}. Specifically, each recording was first denoised (using spectral noise subtraction in Praat \cite{boersma2012praat}) and subsequently level normalized (see also Fig.~\ref{fig:proc_pipeline}). This process led to an improved subjective quality of the recordings and also resulted in higher signal-to-noise ratio (SNR).

\subsubsection{F0, energy, spectral tilt, and duration}

Following pre-processing of the speech data, each recording was initially downsampled to 8 kHz and, subsequently, features were computed using a 25-ms window and 5-ms frame shift. F0 was computed using the YAAPT pitch tracking algorithm \cite{zahorian2008spectral}, signal energy was computed according to Eq. (\ref{eq:energy}) (where x denotes the signal, \textit{t} the current sample, and \textit{w} the frame length; see, e.g., \cite{kakouros20163pro}), and spectral tilt by computing the mel frequency cepstral coefficients (MFCCs) and taking the first (C1) MFCC (see, e.g., \cite{kakouros2018comparison,kakouros2017evaluation}). Word durations were extracted from the corpus annotations. In addition to words, syllable segmentations were estimated from the recordings using a signal envelope-based harmonic oscillator algorithm \cite{rasanen2018pre}. This segmentation method has been shown to compare well with other syllabification methods (see \cite{rasanen2018pre} for more details and comparison). In the present study, the oscillator was set to a centre frequency of 5 Hz with critical damping (Q = 0.5).

\begin{equation} \label{eq:energy}
    EN(t) = \sum\limits_{\tau=-\frac{w}{2}}^{\tau=\frac{w}{2}-1}|x(t+\tau)|^2
\end{equation}

\subsubsection{Normalization}

To account for inter- and intra-talker variation as well as the relationship of some features with their subjective perception, all raw feature values were normalized. Specifically, energy was logarithmically normalized (e.g., \cite{fletcher1933loudness}), F0 was semitone normalized relative to the median F0 for each speaker according to Eq. (\ref{eq:pitch_norm}), spectral tilt was z-score normalized per speaker, and durations (words/syllables) were logarithmically normalized.

\begin{equation} \label{eq:pitch_norm}
    F0'(t) = 12 \cdot \log_2\Big(\frac{F0(t)}{F0_{\mathit{median}}}\Big)
\end{equation}

\subsection{Statistical descriptors}

Five statistical descriptors were computed over words and syllables utilizing the normalized feature values (except for word/syllable duration as descriptors cannot be computed for single values). Specifically, the \textit{mean}, \textit{standard deviation}, \textit{minimum}, \textit{maximum}, and \textit{range} (defined as the difference between the maximum and minimum feature value during a word/syllable) were computed for all data. Words and syllables were selected as the level for analysis as they provide a good basis for the evaluation of potential differences in the data.

\subsection{Supervised classification}

To explore the potential importance of contextual information in the classification of the five dialectal varieties, both sequence classification and context-independent classification methods were utilized. Specifically, three standard context-independent classification algorithms were selected: (i) \textit{k}-nearest neighbours (kNN), (ii) support vector machines (SVM), and (iii) random forests (RF). These algorithms do not make use of the sequence of their inputs. To account for input ordering, as prosodic information is expected to extend across segmental content to temporal segments that spread further than a single segment and might span words and whole utterances, two sequence classifiers were used: (i) linear chain conditional random fields (CRF) and (ii) long short-term memory (LSTM) recurrent neural networks.

The hyperparameters for all classifiers were optimized for best performance in a subset of the test data. Specifically, for the context-independent classifiers and for \textit{k}NN, the number of \textit{k} was set to 10 after heuristically searching for the best \textit{k}, for SVM, a radial basis function (RBF) kernel was used with \textit{C} = 100 and $\sigma$ = 12.0790, and for RF, a classifier with 50 decision trees was utilized. For the sequence classifiers, CRFs were trained using belief propagation and L2 regularization. The maximum number of iterations for CRF training was set to 100 (ensuring training error convergence) and the regularization term to 1. Finally, for the LSTMs, a unidirectional architecture with one hidden LSTM layer containing 128 cells was used and was trained forwards with delays from 0 to 10 frames. Training was carried out with a mini-batch size of 128, 200 epochs, and learning rate of 0.1. At the output layer, a softmax activation function was used to provide the class probability values.

\subsection{Evaluation}

Classification performance was measured both at the word- and syllable-level separately for all classifiers. For each reference unit (word/syllable) the label of the original dialect was compared to the hypotheses provided by the classifiers. Performance was measured in terms of the unweighted average recall (UAR) and classification accuracy.

\section{Results}

\begin{figure}[t]
  \centering
  \includegraphics[width=0.95\linewidth]{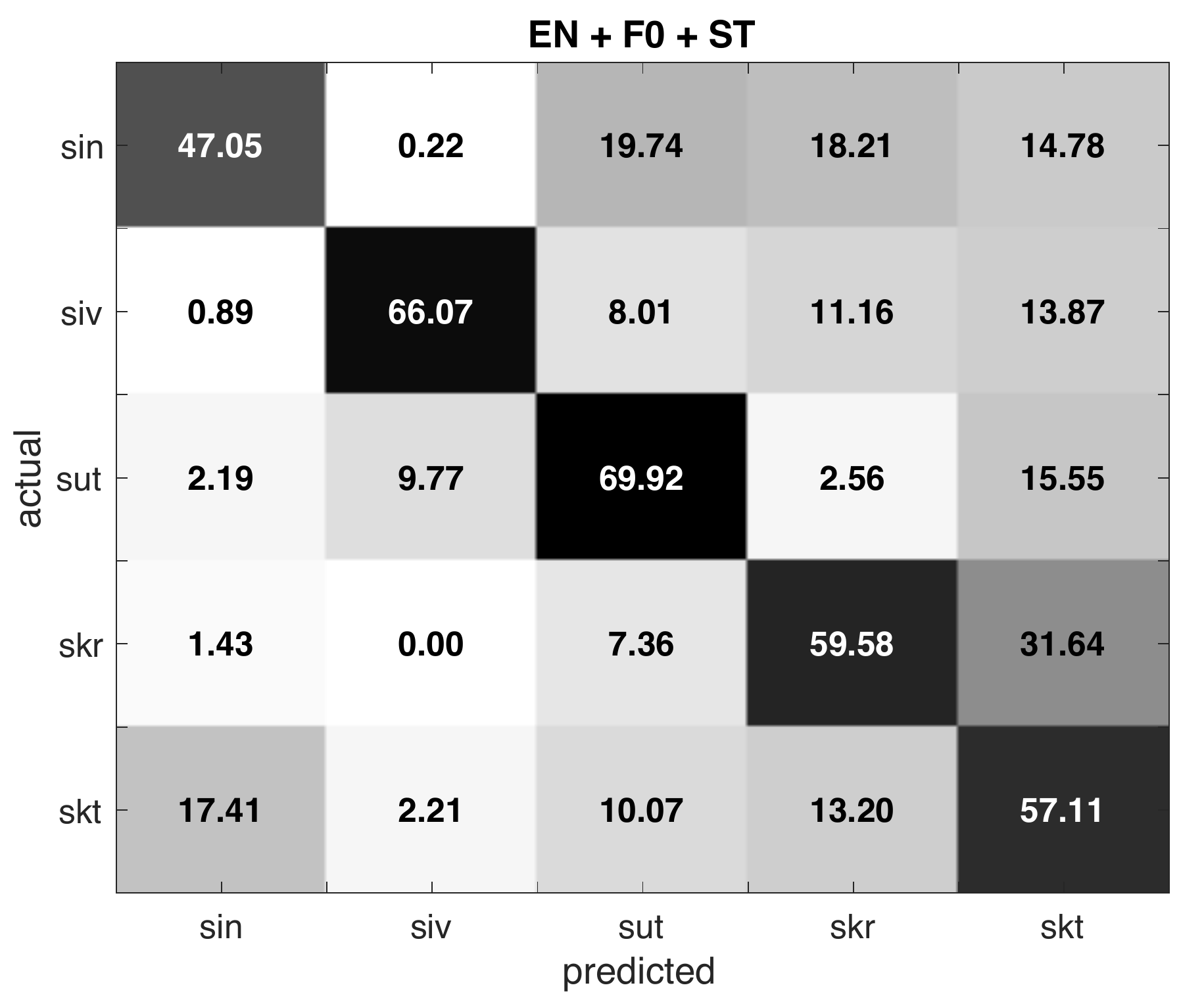}
  \caption{Confusion matrix for the combination of energy, F0, and spectral tilt, over words for the CRF classifier.}
  \label{fig:conf_matrix_crf_nc}
\end{figure}

\begin{figure*}[t]
  \centering
  \includegraphics[width=\linewidth]{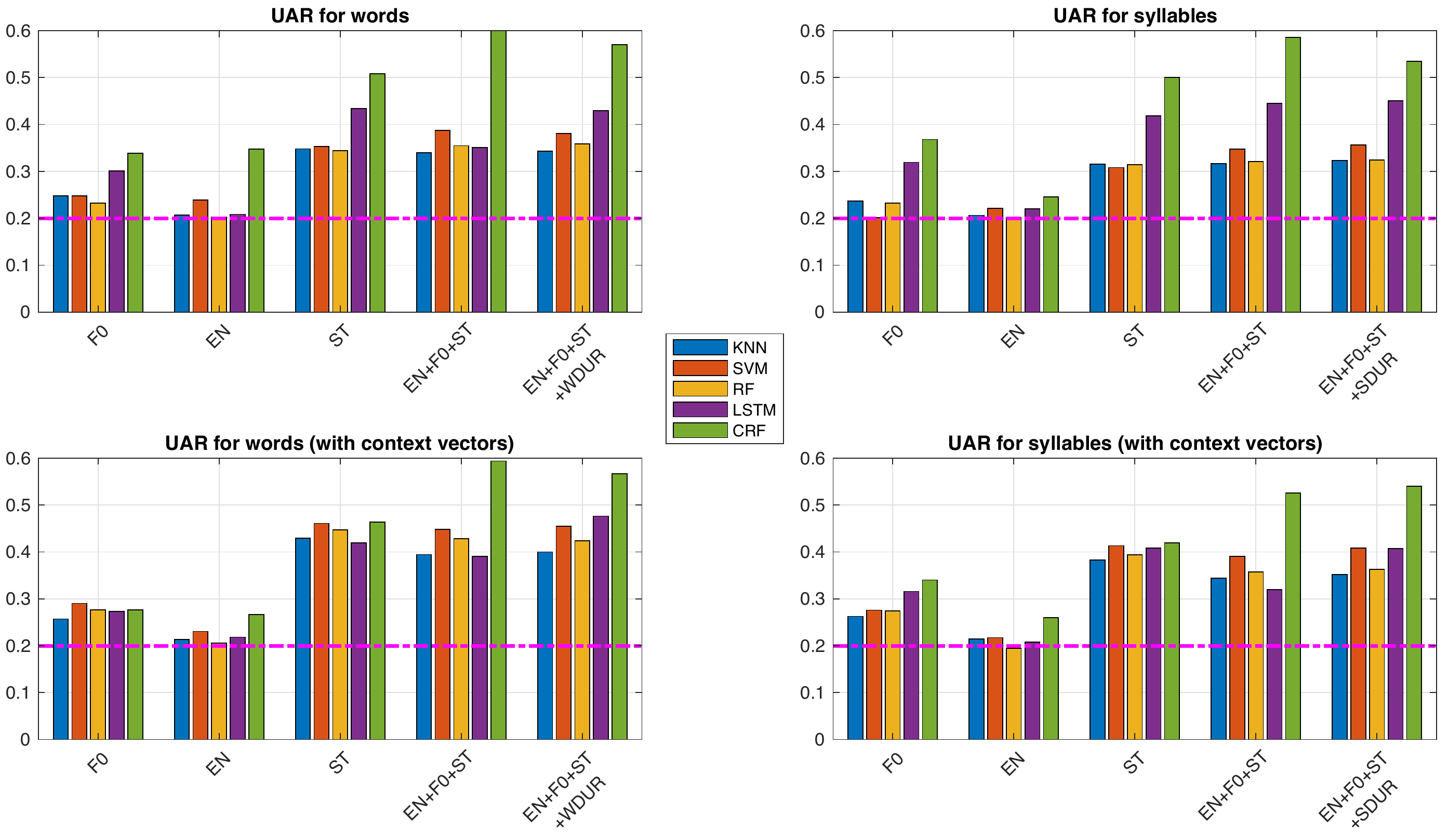}
  \caption{UAR for all classifiers over syllables and words. Horizontal dash dotted line denotes the random baseline performance. Top panel: results with context-independent vectors. Bottom panel: results with context vectors.}
  \label{fig:uar_words_sylls}
\end{figure*}

Experiments were run in a 4-fold evaluation setup (see section~\ref{sec:extendedDigisamiCorpus}) for energy, F0, spectral tilt, duration, and all possible combinations, with features computed over words and syllables separately, and trained using \textit{k}NNs, SVMs, RFs, CRFs, and LSTM networks. A second setup was also tested with the same evaluation procedure, but with the difference that training and testing vectors included the two previous and two forthcoming (word/syllable) vectors to account for contextual dependencies in the data.

Results from the first setup indicate that both the word and syllable context is very important for the classification of the five dialects of S\'{a}mi (see Fig.~\ref{fig:conf_matrix_crf_nc} and Fig.~\ref{fig:uar_words_sylls}). In particular, for the context-independent classifiers (for words), performance for F0 and EN is near the random baseline (20\%), and for ST, it is higher, at approximately 35\%. When context information is introduced with the sequence classifiers, the features of F0 and EN get an increase in performance, both reaching with CRFs a UAR of approximately 34\% and for ST 50.8\%. Similarly, for the overall best feature combination of EN, F0, and ST, context-independent classifiers reached a performance between 35-40\% whereas context-dependent reached up to 60\%. Performance for both words and syllables is similar with the exception that the LSTM network seems to perform better over syllables.

In the second setup, context information was included in the feature vectors. The results provide further support for the importance of acoustic prosodic context in identifying North S\'{a}mi dialects. In this case, the performance improved for the context-independent classification methods, where, for instance, for SVM and words, F0 reached a UAR of around 30\% and ST 46\%. EN did not change in performance and the best overall UAR was 58\% for a combination of EN, F0, and ST with CRFs.

Finally, observing the individual classification performance of the five dialects (Fig.~\ref{fig:conf_matrix_crf_nc}) for the best feature combination of EN, F0, and ST for CRFs, it can be seen that it ranges between 47.05\% and 69.92\%. In particular, the lowest overall performance was for Inari and the highest for the Utsjoki dialect. In general, the performance across the dialects varied with the different classifiers but mostly with the different features and feature combinations utilized. These differences in the performance of prosodic features for the distinct dialects likely reflect dialectal differences in the use of prosodic cues.

\section{Discussion and Conclusions}

The results presented in this work provide evidence that prosodic information is important for the identification of the differences among the five North S\'{a}mi dialects. In particular, from the individual features' perspective, F0, EN, ST, and duration (word/syllable) did not reach high performance when considered independently of their context. However, when context information was introduced, their capacity to discriminate the dialects improved considerably, with the overall best individual feature performance reached for ST (50.8\%) and the best combination for EN, F0, and ST (60\%). Both the syllable- and word-level experiments provided good performance, indicating that in the current task, beyond the selection of the unit of analysis (words/syllables), the inclusion of longer temporal contexts is likely a more important factor. The good performance of ST, a measure that quantifies the relative contribution of high versus low frequency bands of the spectrum\cite{sluijter1996spectral,kakouros2018comparison}, is indicative that ST carries relevant prosodic information for the dialects. However, ST also introduces some segmental information that might have helped in improving performance.

Overall, both the Finnish S\'{a}mi language varieties (69.6\% for Utsjoki, 66.1\% for Ivalo, 47.1\% for Inari) and Norwegian S\'{a}mi (59.6\% for Karasjoki, 57.1\% for Kautokeino) clustered well. It is noteworthy that identifying dialects seems to be a difficult task also for human listeners. For instance, across four regional varieties of Dutch, listeners were able to correctly identify 90\% the country of origin, 60\% the region, and 40\% the province \cite{van1999identification,mccullough2017regional}.

In future work, it will be interesting to extend the experiments to include more sequence classification methods that take into account forward and backward states (bidirectional LSTM networks) and include more data in the evaluations. In addition, beyond the acoustic prosodic investigations, it would be also of interest to include linguistic data in the analysis (phonotactics and lexical analysis).
% Discuss also (i) LSTM performance, (ii) confusion between the different language varieties (Finnish Sami varieties seem to cluster the best with the exception of Inari. Perhaps the lack of sufficient data for Inari (Table 1) does not allow the classifier to learn the prosodic representation). The results of this analysis also agree with earlier investigations on a subset of the data [X]., (iii) the role of tilt. Tilt includes segmental information that might have helped in improving the overall performance.

\section{Acknowledgements}

This study was funded by the Academy of Finland project \textit{Digital Language Typology: Mining from the Surface to the Core} (project no. 12933481). We also thank Kristiina Jokinen and the DigiSami project for the permission to use their speech data.

\newpage
%\pagebreak

\bibliographystyle{IEEEtran}

\bibliography{dlt_refs}

% \begin{thebibliography}{9}
% \bibitem[1]{Davis80-COP}
%   S.\ B.\ Davis and P.\ Mermelstein,
%   ``Comparison of parametric representation for monosyllabic word recognition in continuously spoken sentences,''
%   \textit{IEEE Transactions on Acoustics, Speech and Signal Processing}, vol.~28, no.~4, pp.~357--366, 1980.
% \bibitem[2]{Rabiner89-ATO}
%   L.\ R.\ Rabiner,
%   ``A tutorial on hidden Markov models and selected applications in speech recognition,''
%   \textit{Proceedings of the IEEE}, vol.~77, no.~2, pp.~257-286, 1989.
% \bibitem[3]{Hastie09-TEO}
%   T.\ Hastie, R.\ Tibshirani, and J.\ Friedman,
%   \textit{The Elements of Statistical Learning -- Data Mining, Inference, and Prediction}.
%   New York: Springer, 2009.
% \bibitem[4]{YourName17-XXX}
%   F.\ Lastname1, F.\ Lastname2, and F.\ Lastname3,
%   ``Title of your INTERSPEECH 2019 publication,''
%   in \textit{Interspeech 2019 -- 20\textsuperscript{th} Annual Conference of the International Speech Communication Association, September 15-19, Graz, Austria, Proceedings, Proceedings}, 2019, pp.~100--104.
% \end{thebibliography}

\end{document}